\documentstyle[preprint,aps]{revtex}

\begin{document}
\draft
\preprint{IASSNS-HEP-93/52}
\title{ Algebraic and geometric aspects of generalized quantum
dynamics\\
}
\author{Stephen L. Adler\\}
\address{
Institute for Advanced Study\\
Princeton, NJ 08540\\
}
\author{Yong--Shi Wu\\}
\address{
Department of Physics\\
University of Utah\\
Salt Lake City, UT 84112\\
}
\date{\today}
\maketitle
\begin{abstract}
We briefly discuss some algebraic and geometric aspects
of the generalized Poisson bracket and non--commutative phase
space for generalized quantum dynamics, which are analogous
to properties of the classical Poisson bracket and
ordinary symplectic structure.
\end{abstract}
\pacs{
{\tt$\backslash$\string pacs\{\}}
}
\narrowtext

\section*{}

Recently, one of us (SLA) has proposed a generalization of Heisenberg
picture quantum mechanics, termed {\it generalized quantum dynamics},
which gives a Hamiltonian dynamics for general non--commutative degrees of
freedom.\cite{1,2}  The formalism permits the direct derivation of
equations of motion for field operators, without first proceeding
through the intermediate step of ``quantizing'' a classical theory.  In
a complex Hilbert space, generalized quantum dynamics gives results
compatible with standard canonical quantization.  It is also applicable
to the construction of quantum field theories in quaternionic Hilbert
spaces, where canonical methods fail, basically because the {\it matrix
elements} of operators are themselves elements of the non--commutative
quaternion algebra.  It is hoped that the methods of generalized quantum
dynamics will facilitate answering the question of whether quantum field
theories in quaternionic Hilbert space are relevant to the unification
of the standard model forces with gravitation at energies above the GUT
scale.

As applied to quantum theory, generalized quantum dynamics is formulated
by defining a Hilbert space $V_H$ (based either on complex number or
quaternionic scalars) which is the direct sum of a bosonic space
$V_H^+$ and a fermionic space $V_H^-$.  Next, following Witten\cite{3}, one
defines an operator $(-1)^F$ with eigenvalue $+1$ for states in
$V_H^+$ and $-1$ for states in $V_H^-$.
Finally, one needs a trace operation ${\bf Tr}\,{\cal O}$ for a
general operator ${\cal O}$, defined by
\begin{equation}
{\bf Tr}~{\cal O}= {\rm
Re}\,Tr\,(-1)^F{\cal O}= {\rm Re} \sum_n~\langle n|(-1)^F{\cal
O}|n\rangle\label{one}.
\end{equation}
It is easy to show that the trace ${\bf Tr}$ vanishes for
operators ${\cal O}$ which anti--commute with $(-1)^F$, and so ${\bf
Tr}~{\cal O}$ acts non--trivially only on the part of ${\cal O}$ which
commutes with $(-1)^F$.

Let $\{q_r(t)\}$ be a finite set of time--dependent quantum variables, which
act as operators on the underlying Hilbert space, with each individual
$q_r$ of either bosonic or fermionic type, defined respectively as
commuting, or anti--commuting with $(-1)^F$.  No other {\it a priori}
assumptions about commutativity of the $q_r$ are made.  The
Lagrangian ${\bf L}[\{q_r\},\{\dot q_r\}]$ is then defined as the trace of a
polynomial function of $\{q_r(t)\}$ and its time derivative $\{\dot
q_r(t)\}$, or as a suitable limit of such functions.  The action
${\bf S}$ is defined as the time integral
of ${\bf L}$, and generalizations of the Euler--Lagrange equations
follow from the requirement that $\delta {\bf S} = 0$ for arbitrary
(same--type) variations of the operators.  Derivatives of ${\bf L}$ with
respect to $q_r$ and $\dot q_r$ are defined by writing the variation
of ${\bf L}$, for infinitesimal variations in the $\{q_r\}$, in the form,
\begin{equation}
\delta {\bf L} = {\bf Tr}~\sum_r~\left({\delta{\bf L}\over\delta q_r}~\delta
q_r
+{\delta{\bf L}\over\delta\dot q_r}~\delta\dot q_r\right)\label{two},
\end{equation}
where cyclic permutations of operators inside ${\bf Tr}$ have been used
to order $\delta q_r$ and $\delta\dot q_r$ to the right.  The momentum
$p_r$ conjugate to $q_r$ is defined by
\begin{equation}
{{\delta {\bf L}}\over{\delta \dot q_r}}=p_r\label{three},
\end{equation}
and the Hamiltonian ${\bf H}$ is given by
\begin{equation}
{\bf H}={\bf Tr}~\sum_r~p_r\dot q_r-{\bf L}\label{four}.
\end{equation}

In complete analogy with the Lagrangian derivatives defined in Eq. (2),
for a general trace functional
${\bf A}$, constructed as the trace ${\bf Tr}$ of a (bosonic)
polynomial function of operator arguments, one can define a unique
derivative $\delta {\bf A}/ \delta q_r$ with respect to
the operator $q_r$ (and of the same
bosonic or fermionic type as $q_r$) by the relation
\begin{equation}
\delta {\bf A}={\bf Tr}~{\delta{\bf A}\over \delta q_r}~\delta
q_r\label{five}.
\end{equation}
Again, cyclic invariance of the trace has been used to reorder all
$\delta q_r$ factors to the right in the respective terms in which they
occur.  Using this derivative, one can then define a generalized Poisson
bracket, as follows.  Let $\{q_r\},~\{p_r\}$ be the set of operator phase
space variables introduced above, which for each $r$ are either
both bosonic or both fermionic, in the sense that they commute or
anticommute with $(-1)^F$.  Again, no further
{\it a priori} assumptions are made about their commutativity.  If we
now let ${\bf A}[\{q_r\},~\{p_r\}]$ and ${\bf B}[\{q_r\},~\{p_r\}]$ be
two trace functionals of their arguments, then the generalized Poisson
bracket $\{{\bf A},~{\bf B}\}$ is defined by
\begin{equation}
\{ {\bf A},~{\bf B}\}={\bf Tr}
\left[\sum_r~\varepsilon_r\left({\delta{\bf A}\over\delta
q_r}~{\delta{\bf B}\over\delta p_r}-{\delta{\bf B}\over\delta
q_r}~{\delta{\bf A}\over\delta p_r}\right)\right]\label{six},
\end{equation}
with $\varepsilon_r=+1 (-1)$ according as whether $q_r$ and $p_r$ are
bosonic (fermionic).  Using the generalized bracket, the time
development of a general trace functional ${\bf A}[\{q_r\},~\{p_r\},~t]$
takes the form\cite{1,2}
\begin{equation}
{d{\bf A}\over dt}={\partial {\bf A}\over \partial t}
+\{{\bf A},~{\bf H}\}\label{seven},
\end{equation}
with ${\bf H}$ the total trace Hamiltonian.  It was conjectured in
Refs. 1 and 2 that the generalized bracket obeys the Jacobi identity,
\begin{equation}
0=\{{\bf A},~\{{\bf B},{\bf C}\}\}+\{{\bf C},~\{{\bf A},~{\bf B}\}\}+
\{{\bf B},~\{{\bf C},~{\bf A}\}\}\label{eight},
\end{equation}
and this conjecture has recently been proved by Adler, Bhanot, and
Weckel.\cite{4}  The key observation is that despite the absence of
both commutativity and the product rule, and the lack of a definition
for the double derivative, pair--wise cancellations still occur in
the right--hand side of Eq. (8) because of cyclic permutability
inside the trace ${\bf Tr}$.  The proof of Eq. (8) is in fact independent of
the Hilbert space arena on which the operators $\{q_r\}$ act.  All that
is used are the definition of derivative of Eq. (5), and the assumptions
that operator
multiplication is associative, and that there exists a graded
trace ${\bf Tr}$ permitting cyclic permutation of
non--commuting operator variables, according to the formula
\begin{equation}
{\bf Tr}~{\cal O}_{(1)}{\cal O}_{(2)}=\pm{\bf Tr}~{\cal O}_{(2)}{\cal
O}_{(1)}\label{nine},
\end{equation}
with the $+~(-)$ sign holding when ${\cal O}_{(1)}$ and ${\cal O}_{(2)}$
are both bosonic (fermionic).

Evidently the generalized bracket of Eq. (6) can be viewed
as an extension of the classical Poisson bracket, which permits
the introduction of non--commuting phase space variables
$\{q_r\},~\{p_r\}$. Our aim in this note is to document a number
of further algebraic and geometric properties of non--commutative
phase space, which closely relate to the existence of the
generalized Poisson bracket that satisfies the Jacobi identity
of Eq. (8), but which do not enter into the proof
given in Ref. 4.

The first of these involves the algebraic structure of the trace
functionals, under the product operation used to construct the
antisymmetric bracket
of Eq.~(6).  Letting ${\bf A}$ and ${\bf B}$ be any two trace
functionals defined on phase space, a product ${\bf A}\circ {\bf B}$
that remains a trace functional can be defined by
\begin{equation}
{\bf A}\circ {\bf B}\equiv{\bf Tr}\left[\sum_r~\varepsilon_r{\delta{\bf
A}\over \delta q_r}~{\delta{\bf B}\over\delta p_r}\right]\label{ten},
\end{equation}
in terms of which the generalized Poisson bracket takes the form of a
commutator,
\begin{equation}
\{{\bf A}, {\bf B}\}={\bf A}\circ {\bf B}-{\bf B}\circ {\bf
A}\label{eleven}.
\end{equation}
The algebra ${\cal A}_\circ$ of trace functionals under the product
$\circ$ can now be characterized in terms of the standard
classification\cite{5} of non--associative algebras.  It is {\it associative}
iff the associator $({\bf A},{\bf B},{\bf C})$ defined by
\begin{equation}
({\bf A}, {\bf B}, {\bf C})\equiv ({\bf A}\circ {\bf B})\circ {\bf C}-
{\bf A}\circ ({\bf B}\circ {\bf C})\label{twelve}
\end{equation}
vanishes.  It is {\it flexible} iff the associator obeys
\begin{equation}
({\bf A}, {\bf B}, {\bf C})=-({\bf C}, {\bf B}, {\bf A})\label{thirteen},
\end{equation}
and it is {\it Lie--admissible} iff the associator obeys
\begin{eqnarray}
0=({\bf A}, {\bf B}, {\bf C})-({\bf A}, {\bf C}, {\bf B})+
({\bf B}, {\bf C}, {\bf A})\nonumber\\
                             -({\bf B},{\bf A}, {\bf C})+
({\bf C},{\bf A}, {\bf B})-({\bf C},{\bf B},{\bf A})\label{fourteen}.
\end{eqnarray}
Evidently, any associative algebra is Lie--admissible, but the converse is
of course not true.  Now by substituting Eq. (12) into Eq. (14) and
rearranging using Eq. (11), we find that Eq. (14) is equivalent to
\begin{equation}
0=\{{\bf A}, \{{\bf B}, {\bf C}\}\}+\{{\bf C},\{{\bf A}, {\bf B}\}\}+
\{{\bf B}, \{{\bf C}, {\bf A}\}\}\label{fifteen},
\end{equation}
which is true by virtue of the Jacobi identity for the generalized
Poisson bracket.  To see that Eq. (12) does not vanish and that Eq. (13)
does not hold, it
suffices to consider the special case in which the variables $\{q_r\}$
and $\{p_r\}$ are commuting (bosonic) $c$--numbers.  This is just the
classical case in which $\{{\bf A}, {\bf B}\}$ is proportional to the
standard Poisson bracket, and a simple calculation of multiple
derivatives [see, e.g., Ref. 5, Sec. 7.3] shows that both the vanishing
of Eq. (12) and the identity of
Eq. (13) are false for the product defined by Eq. (10).  Hence the
algebra ${\cal A}_\circ$ is neither associative nor flexible,
and therefore is only of secondary interest.  But as in the case of
its classical analog, ${\cal A}_\circ$ is Lie--admissible
by virtue of the Jacobi identity, and hence the resulting Lie structure
defined by Eq. (11) is of primary importance.  Thus, the trace functionals
form a Lie algebra under the generalized Possion bracket of Eq. (11)
and, in particular, the total trace conserved symmetry generators
that commute with the total trace Hamiltonian form a Lie
subalgebra.\cite{4}

The second aspect to be discussed relates to the tangent vector
fields associated with the generalized dynamics. Let $X_{\bf A}$
be the tangent vector field associated with a trace functional
${\bf A}$, defined as a formal derivative operator by
\begin{equation}
X_{\bf A} \equiv {\bf Tr}~\left[\sum_r~\left(\varepsilon_r{\delta{\bf
A}\over\delta q_r}~{\delta\over\delta p_r}-{\delta{\bf
A}\over\delta p_r}~{\delta\over\delta q_r}\right)\right]\label{sixteen},
\end{equation}
and defined operationally by its action on any trace functional ${\bf B}$,
\begin{equation}
X_{\bf A}{\bf B} = {\bf B}X_{\bf A} + (X_{\bf A}{\bf B})\label{seventeen},
\end{equation}
with $(X_{\bf A}{\bf B})$ given by
\begin{eqnarray}
(X_{\bf A}{\bf B})
={\bf Tr}\left[\sum_r~\left(\varepsilon_r
{\delta{\bf A}\over\delta q_r}~{\delta{\bf B}\over\delta p_r}-
{\delta{\bf A}\over\delta p_r}~{\delta{\bf B}\over\delta
q_r}\right)\right]\nonumber\\
={\bf Tr}\left[\sum_r~\varepsilon_r\left(
{\delta{\bf A}\over\delta q_r}~{\delta{\bf B}\over\delta p_r}-
{\delta{\bf B}\over\delta q_r}~{\delta{\bf A}\over\delta
p_r}\right)\right]=\{{\bf A},{\bf B}\}.\label{eighteen}
\end{eqnarray}
In terms of this operator, the time development of a general
trace functional ${\bf B}[\{q_r\},~\{p_r\}]$, under the dynamics
governed by ${\bf A}$ as total trace Hamiltonian, can be
rewritten as [cf Eq.~(7)]
\begin{equation}
{d{\bf B}\over dt}= - (X_{\bf A}{\bf B})\label{nineteen}.
\end{equation}
Thus the tangent vector field $X_{\bf A}$ can be viewed as (minus) the
directional derivative along the time evolution orbit (called the
phase flow in Ref. 6) of the point $(\{q_r\}, \{p_r\})$ in phase space,
which is determined by the Hamiltonian equations of motion\cite{1}
\begin{equation}
{d{q_r}\over dt}= \varepsilon_r {\delta {\bf A}\over
\delta p_r}~, ~~~~~~~~
{d{p_r}\over dt}= - {\delta {\bf A}\over \delta q_r}\label{twenty},
\end{equation}
with ${\bf A}$ acting as the total trace Hamiltonian. Following Ref. 6,
we call a tangent vector field of the form of Eq. (16) a
Hamiltonian vector field, the same name as for its classical
counterpart.

We note that with respect to the product defined by Eq. (10), the
directional derivative $X_{\bf A}$ does not obey the
Leibniz product rule,
\begin{equation}
(X_{\bf A} ({\bf B}\circ{\bf C}))
\neq (X_{\bf A}{\bf B})\circ {\bf C}
+ {\bf B}\circ (X_{\bf A}{\bf C})\label{twenty-one}.
\end{equation}
(It is easy to verify that the same is true in the
classical case.) However, it does obey the Leibniz
product rule for the generalized Poisson bracket
or the commutator defined by Eq. (11),
\begin{equation}
(X_{\bf A} \{{\bf B}, {\bf C}\})
= \{(X_{\bf A}{\bf B}), {\bf C}\}
+ \{{\bf B}, (X_{\bf A}{\bf C})\}\label{twenty-two},
\end{equation}
because, in view of Eq. (14), this equation is equivalent to
the Jacobi identity of Eq. (8).

What is the algebraic structure of the Hamiltonian vector fields?
Let us compute the action of the commutator of two tangent vector
fields $X_{\bf A}$ and $X_{\bf B}$ on a third trace
functional ${\bf C}$,
\begin{eqnarray}
([X_{\bf A},~X_{\bf B}]{\bf C})
&=&(X_{\bf A}(X_{\bf B}{\bf C}))-(X_{\bf B}(X_{\bf A}{\bf C}))\nonumber\\
&=&\{{\bf A},\{{\bf B},{\bf C}\}\}-\{{\bf B},\{{\bf A},{\bf
C}\}\}\nonumber\\
&=&\{{\bf A},\{{\bf B},{\bf C}\}\}
+\{{\bf B},\{{\bf C},{\bf A}\}\}\label{twenty-three}.
\end{eqnarray}
Using Eq. (14) with ${\bf A}$ replaced by $\{{\bf A},
{\bf B}\}$ and ${\bf B}$ replaced by ${\bf C}$, we also get
\begin{equation}
(X_{\{{\bf A},{\bf B}\}} {\bf C})=\{\{{\bf A}, {\bf B}\},
{\bf C}\}\label{twenty-four},
\end{equation}
and subtracting Eq. (24) from Eq. (23) gives finally
\begin{eqnarray}
&&\left(([X_{\bf A},X_{\bf B}]-X_{\{{\bf A}, {\bf B}\}}){\bf
C}\right)\nonumber\\
&=&\left\{{\bf A}, \{{\bf B}, {\bf C}\}\right\}+\left\{{\bf B},\{{\bf C}, {\bf
A}\}\right\}+
\left\{{\bf C}, \{{\bf A}, {\bf B}\}\right\}\nonumber\\
&=&0\label{twenty-five}.
\end{eqnarray}
Hence validity of the Jacobi identity for the generalized Poisson
bracket implies that the Hamiltonian vector fields $X_{\bf A}$
defined by Eqs. (16)--(18) obey the commutator algebra
\begin{equation}
[X_{\bf A}, X_{\bf B}] = X_{\{{\bf A}, {\bf B}\}}\label{twenty-six},
\end{equation}
and, therefore, form a Lie algebra that is isomorphic to the
Lie algebra of trace functionals under the generalized Poisson
bracket, which is the generalized quantum dynamics analog of
a standard result\cite{6} in classical mechanics.

Finally, we address the geometric structure underlying
generalized quantum dynamics. As is well--known, there is
a geometry which underlies classical Hamiltonian dynamics,
namely the symplectic geometry of ordinary phase space.  Can
we generalize symplectic geometry to
non--commutative phase space?  If a generalized
symplectic structure exists, is it preserved by phase space flows
(or Hamiltonian time evolutions) as in classical mechanics?\cite{6}
In the following we present a discussion of these questions
with affirmative answers, which is readable to physicists who
are not familiar with differential forms.\cite{7}

Ordinary symplectic geometry is
defined by a standard (constant) anti--symmetric metric in the
tangent or cotangent spaces of a phase space.
(By way of contrast, Riemannian geometry, which is perhaps more familiar
to physicists,
is defined by a {\it symmetric} metric in the tangent or cotangent
spaces of a manifold.)  To avoid
differential forms, let us consider the cotangent space,
which is known to be spanned by covariant vectors whose
components form the gradient (or differential) of a function
on phase space. The standard (anti--symmetric) symplectic
metric, or the inner product, between two covariant vectors
that are the gradients of two classical functions $A(q_r, p_r)$
and $B(q_r, p_r)$ on phase space, is provided by the
classical Poisson bracket $\{A, B\}$. In a non--commutative
phase space, the analogs of functions are trace
functionals, and the analogs of the differentials of functions
are the differentials of trace functionals, i.e., Eq. (5)
adapted to phase space,
\begin{equation}
\delta {\bf A}={\bf Tr}~\left[\sum_r~\left(
{\delta{\bf A}\over \delta q_r}~\delta q_r~
+ {\delta{\bf A}\over \delta p_r}~\delta
p_r\right)\right]\label{twenty-seven}.
\end{equation}
With the generalized Poisson bracket of Eq. (6) available,
we can use it to define a generalized symplectic structure
$\Omega$ on the non--commutative phase space,
through defining the inner product between two cotangent
vectors $\delta {\bf A}$ and $\delta {\bf B}$ as follows,
\begin{eqnarray}
\Omega(\delta {\bf A}, \delta {\bf B})
&=&\{{\bf A},~{\bf B}\}\nonumber\\
&\equiv& {\bf Tr} \left[\sum_r~\varepsilon_r\left(
{\delta{\bf A}\over\delta q_r}~{\delta{\bf B}
\over\delta p_r}-{\delta{\bf B}\over\delta
q_r}~{\delta{\bf A}\over\delta p_r}\right)\right]\label{twenty-eight}.
\end{eqnarray}

To see that such a symplectic structure is preserved by
any Hamiltonian phase flow of Eq. (20), we observe that
the time derivative of the inner product along the
phase--flow orbit is
\begin{equation}
{d\over dt}~ \Omega(\delta {\bf B}, \delta {\bf C})
= {d\over dt}~ \{{\bf B},~{\bf C}\}
= \left\{ \{{\bf B},~{\bf C}\},~{\bf A}\right\}\label{twenty-nine},
\end{equation}
while that of the differential $\delta {\bf B}$ along
the same flow is
\begin{equation}
{d \over dt}~ \delta {\bf B} \equiv \delta {\dot{\bf
B}}~\label{thirty}
\end{equation}
where the dot abbreviates the time--derivative.  Therefore,
we have
\begin{eqnarray}
\Omega (\delta {\dot{{\bf B}}},~\delta{\bf C})
+\Omega (\delta{\bf B},~\delta {\dot{{\bf C}}})
= \{ {\dot{{\bf B}}},~{\bf C}\}
+ \{{\bf B},~{\dot{{\bf C}}}\}\nonumber\\
= \{\{{\bf B},~{\bf A}\}, {\bf C}\}
+ \{{\bf B},~\{{\bf C}, {\bf A}\}\}~.\label{thirty-one}
\end{eqnarray}
Therefore the Jacobi identity of Eq. (8) implies
\begin{equation}
{d\over dt}~ \Omega(\delta {\bf B}, \delta {\bf C})
= \Omega (\delta {\dot{{\bf B}}},~\delta {\bf C})
+\Omega (\delta {\bf B},~\delta {\dot{{\bf C}}})\label{thirty-two},
\end{equation}
that is, the symplectic
structure is invariant under Hamiltonian phase flow.
This statement can be viewed as a (dual) form of the generalized
quantum dynamics analog of the Liouville theorem.

Thus, generalized quantum dynamics, albeit with non--commuting
operator phase space variables, has an underlying generalized
symplectic geometry which is preserved by the time evolution generated
by
any total trace Hamiltonian. Basically this is due to the
existence of a (graded) trace ${\bf Tr}$ that permits
cyclic permutation of non--commuting operator variables,
which implies the validity of the Jacobi identity for the
generalized Poisson bracket. As in classical mechanics,
we expect that the basic concepts and theorems
of generalized quantum dynamics will be invariant
under the group of symplectic transformations,
i.e., under transformations which preserve the
generalized symplectic structure.

To conclude, we have seen that in many algebraic and
geometric aspects, the generalized quantum dynamics proposed
in Refs. 1 and 2 is analogous to classical mechanics. It is
really surprising that with the help of a cyclically permutable
(graded) trace alone, so many features of classical mechanics
can be generalized to a non--commutative phase space. (We remind
readers once more that in Ref. 1 and in our present discussion,
{\it no phase space variable commutation relations such
as commutativity, anti--commutativity, or
$q$--commutators are assumed}.)  Further developments in
generalized quantum dynamics, paralleling to some extent aspects of
existing quantization schemes, are expected.

\acknowledgments

The authors wish to thank F.J. Dyson for a discussion with SLA,
and acknowledge the hospitality of the Aspen Center for Physics,
where this work was done.  SLA was supported in part by the
Department of Energy under Grant No. DE--FG02--90ER40542.
Y--SW was supported in part by the National Science Foundation under Grant
No. PHY--9309458.


\begin{references}
\bibitem{1}
S.L. Adler, {\it Generalized quantum dynamics}, Nucl. Phys. B (in press).
\bibitem{2}
S.L. Adler, {\it Quaternionic Quantum Mechanics and Quantum Fields},
Oxford University Press, to be published in 1994.
\bibitem{3}
E. Witten, J. Diff. Geom. {\bf 17} (1982) 661.
\bibitem{4}
S.L. Adler, G.V. Bhanot, and J.D. Weckel, {\it Proof of Jacobi identity
in generalized quantum dynamics}, J. Math. Phys. (in press).
\bibitem{5}
S. Okubo, {\it Introduction of Octonion and other Non--Associative
Algebras in Physics}, Cambridge University Press, to be published.
\bibitem{6}
V.I. Arnold, {\it Mathematical Methods of Classical Mechanics}, p. 211,
Springer--Verlag, New York, 1978; R. Abraham and J.E. Marsden, {\it
Foundations of Mechanics}, second edition, p. 194, Benjamin/Cummings,
Reading, MA, 1980.
\bibitem{7}
A more mathematical treatment, using the notions of generalized
differential forms in non--commutative phase space, may be possible.
This is left for a future publication.
\end{references}
\end{document}